\documentclass[10pt,letterpaper]{article}
\usepackage{opex3}

\usepackage{amsmath}
\usepackage{graphicx,dcolumn,bm}
\usepackage{mathrsfs}
\usepackage[usenames,dvipsnames]{xcolor}
\usepackage[colorlinks=true,citecolor=MidnightBlue,linkcolor=MidnightBlue,urlcolor=MidnightBlue]{hyperref}
\usepackage{braket}

\definecolor{blue1}{rgb}{0.72,0.82,0.98}
\definecolor{silver}{rgb}{0.6,0.6,0.6}
\definecolor{silver1}{rgb}{0.8,0.8,0.8}

\newcommand{\operator}[1]{\hat{#1}}
\renewcommand{\vector}[1]{\mathbf{#1}}

\begin{document}

\title{Nonlocal photon correlations and violation of Bell inequalities for spatially separated classical light fields}

\author{D. Bhatti$^{1}$, R. Schneider$^{1,2}$, T. Mehringer$^{1,2}$, S. Oppel$^{1}$,\\ J. von Zanthier$^{1,2,*}$}

\address{$^1$Institut f\"ur Optik, Information und Photonik, Universit\"at Erlangen-N\"urnberg, 91058 Erlangen, Germany\\
$^2$Erlangen Graduate School in Advanced Optical Technologies (SAOT), Universit\"at Erlangen-N\"urnberg, 91052 Erlangen, Germany}

\email{$^*$Joachim.vonZanthier@physik.uni-erlangen.de} 

\begin{abstract}
It is theoretically and experimentally shown that photons emitted by statistically independent incoherent classical light sources and measured in the far field in spatially separated modes may display spatial correlations akin to path-entanglement of photons produced by quantum sources. By measuring higher order photon-correlations at different locations, i.e.,  $m$ photons in one mode and one photon in another mode, we experimentally demonstrate for $m \geq 6$ a violation of Bell-type inequalities for spatial degrees of freedom.  The spatial correlations among the photons can be understood from state projection where the detection of the first $m$ photons projects the sources onto a state which emits the subsequent photon in a strongly correlated manner. 
From this perspective the entanglement and violation of Bell's inequalities appears as a consequence of 
nonvanishing cross correlations between noncommuting quadrature phase components of the two spatially 
separated fields after $m$ photons have been recorded. 
In this way we show that classical systems may produce spatial field correlations violating local realistic theories.
\end{abstract}


\section{Introduction}

Since Bohr formulated his correspondence principle \cite{Bohr1920} it is arguably agreed that classical physics can be used to explain quantum phenomena 
in the classical limit, i.e., in the case of high quantum numbers. For example, electron wavefunctions reduce to Bohr orbits in the limit of high principal and orbital quantum numbers, and laser amplitudes display vanishing fluctuations in the limit of an infinite mean photon flux.
By contrast, it is less anticipated that quantum mechanical predictions, i.e., implications considered to be purely quantum mechanical, can be fruitfully applied for a better understanding of classical systems. 
In this context it was recently proposed that a violation of Bell's inequalities, so far considered to be a typical quantum mechanical phenomenon displayed only by quantum systems, may  be used to quantify  correlations between coupled degrees of freedom of classical systems 
\cite{Spreeuw98,Aiello05,Eberly11,Simon10,Khoury10,Leuchs11,Saleh13,Agarwal13,Eberly14}. Such strong correlations of classical systems, called \textit{classical entanglement} or \textit{nonquantum entanglement} in the literature \cite{Simon10,Saleh13}, have been used to describe the specific coherence properties of classical light beams,
either for discrete or continuous variables \cite{Wolf03,Simon10,Leuchs11,Saleh13,Agarwal13}. However, classical entanglement has been demonstrated so far only for coupled degrees of freedom of the \textit{same particle}, e.g., amplitude and polarization of a photon in a classical light beam,  and was expected to reach its limits if true nonlocal multiparticle entanglement is considered. It has thus been conjectured that nonlocal \textit{multiparticle} entanglement is of exclusive quantum nature \cite{Spreeuw98}. 
In this letter we demonstrate that Bell's inequalities can be equally violated with spatially separated particles produced by classical systems, i.e., photons emitted by classical light sources recorded in spatially separated modes. We thus show that nonlocal multiparticle entanglement is not restricted to the quantum world but may be observed also with classical systems. 

The paper is organized as follows: In Sect.~2 we present the system under investigation and introduce the normalized first and second order spatial intensity correlation functions employed for Bell's inequalities involving spatial degrees of freedom. In Sect.~3 we expand the analysis and introduce intensity correlation functions of higher order which we use, with a new interpretation, to violate the Bell inequalities. In Sect.~4  we present our experimental results and in Sect.~5 we finally conclude.

\section{System under investigation}
Recently, we demonstrated that photons emitted by statistically independent quantum sources, e.g., single photon emitters (SPE), may be path-entangled if the photons are recorded in the far field such that the individual photon source remains unknown \cite{Wiegner2010}. The entanglement was proven by a violation of Bell's inequalities \cite{Bell64,Clauser74} formulated for spatial degrees of freedom. 
The question remains whether such nonlocal correlations among photons recorded in spatially separated  modes can be produced with classical light, i.e., photons emitted by classical sources. To investigate this question we generalize in what follows the system of two SPE discussed in \cite{Wiegner2010} to classical light sources. 

\begin{figure}
	\centering
\includegraphics[width=0.6\textwidth]{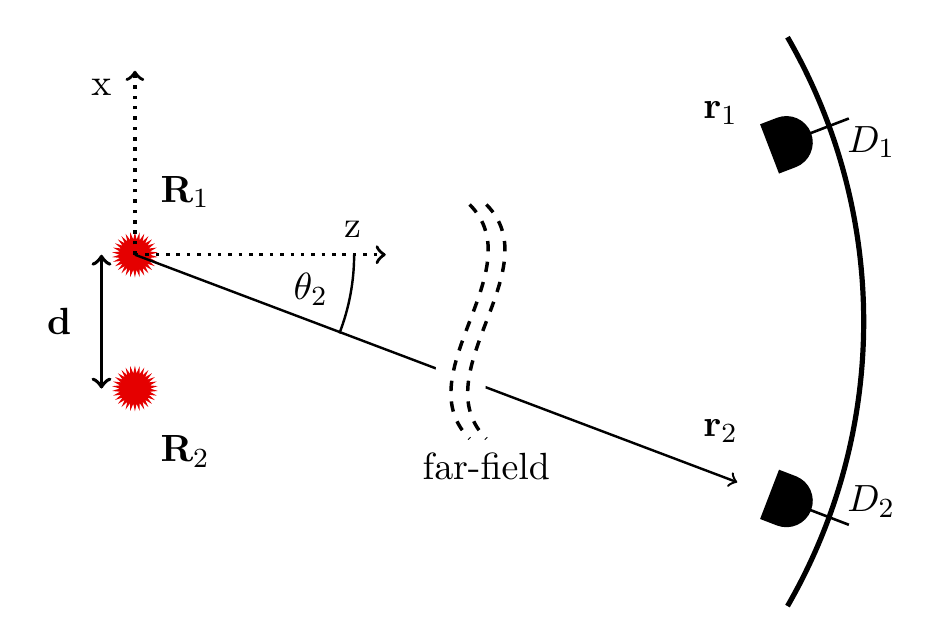}
\caption{Scheme of the investigated setup. Two statistically independent Thermal Light Sources (TLS) at $\vector{R}_{1}$ and $\vector{R}_{2}$, separated by a distance $d$, emit photons which are registered by two detectors $D_1$ and $D_2$ in the far-field at $\vector{r}_{1}$ and $\vector{r}_{2}$, respectively. Assuming that each detector registers coincidentally exactly one photon the second order correlation function  
$G^{(2)}_{2 \ TLS}(\vector{r}_{1}, \vector{r}_{2})$ is measured.}
\label{fig:detectionscheme}
\end{figure}

We thus consider a system of two statistically independent identical classical light sources, e.g., thermal light sourcess (TLS), located at $\vector{R}_{l}$, $l = 1, 2$, and two detectors placed in the far-field of the sources at $\vector{r}_{j}$, $j = 1, 2$ (see Fig.~\ref{fig:detectionscheme}). Due to the far field condition and the indistinguishability of the emitted photons the electric field operator $\operator{E}^{(+)}(\vector{r}_{j})$ is written in the form
\begin{equation}
\label{E-field}
	\operator{E}^{(+)}(\vector{r}_{j})=\mathcal{E}_{0} \ \left( \hat{a}_{1} + e^{i \delta_{j}} \hat{a}_{2} \right) = \left( \operator{E}^{(-)}(\vector{r}_{j}) \right)^\dagger\ ,
\end{equation}
where $\mathcal{E}_{0}$ is the amplitude of the total electric field at $\vector{r}_{j}$ and $\delta_{j}=\delta_{j}(\vector{r}_{j})= k\frac{\vector{r}_{j} \cdot \vector{R}_{2}}{|\vector{r}_{j}|}= k d \ \sin(\theta_{j})$ describes the relative optical phase accumulated by a photon emitted from the source at $\vector{R}_{2}$ with respect to a photon emitted from the source at $\vector{R}_{1}$, both reaching the $j$-th detector at $\vector{r}_{j}$ (cf. Fig.~\ref{fig:detectionscheme}). Here, $k=\frac{2\pi}{\lambda}$ is the wave number of the photons, the operator $\hat{a}_{l}$ denotes the annihilation of a photon from source $l$ and the electric field is described by a scalar since we only consider a single polarization. 

Starting from the Bell inequalities for continuous variables \cite{Clauser74}
\begin{equation}
	-XY\leq xy-xy'+x'y+x'y'-x'Y-yX \leq 0 \; ,
\label{eq:ch74}
\end{equation}
which hold for all $x, \ x'\leq X$ and  $y, \ y'\leq Y$, we can, by setting $X = Y = 1$, interpret $x, x', y,  y'$ as probabilities and use  Glauber's first and second order intensity correlation functions \cite{Glauber63} as input to Eq.~(\ref{eq:ch74}). Hitherto the spatial intensity correlation functions of first and second order at equal times, defined as \cite{Glauber63}
\begin{equation}
\begin{split}
G^{(1)}(\vector{r}_{1}) = &\left< \operator{E}^{-}(\vector{r}_{1}) \operator{E}^{+}(\vector{r}_{1}) \right> \\
G^{(2)}(\vector{r}_{1},\vector{r}_{2}) = & \left< \operator{E}^{-}(\vector{r}_{1}) \operator{E}^{-}(\vector{r}_{2}) \operator{E}^{+}(\vector{r}_{2}) \operator{E}^{+}(\vector{r}_{1}) \right> \; ,
\end{split}
\end{equation}
have to be cast into a form as to be interpreted as detection probabilities of single photon detection events $P^{(1)}(\vector{r}_{1})\propto G^{(1)}(\vector{r}_{1})$ and joint two-photon detection events $P^{(2)}(\vector{r}_{1},\vector{r}_{2})\propto G^{(2)}(\vector{r}_{1},\vector{r}_{2})$, respectively \cite{Wiegner2010}.

If both light sources are incoherent and statistically independent the emitted intensity is distributed isotropically in space. The first order correlation function $G^{(1)}(\vector{r}_{1})$ is thus a constant, independent of $\vector{r}_{1}$.
By contrast, the second order correlation function for two statistically independent SPE reads \cite{Scully97}
\begin{equation}
\begin{split}
	G^{(2)}_{2 \ SPE}(\vector{r}_{1},\vector{r}_{2})
	= 2 \ \mathcal{E}_{0}^{4} \ \left(1 + \cos(\delta_{1}-\delta_{2}) \right) \ ,
\label{eq:G2SPE}
\end{split}
\end{equation}
whereas for two statistically independent TLS with identical mean photon numbers $\left<\hat{n}_{j}\right>=\left<\hat{n}\right>$ it is given by  \cite{Scully97}
\begin{equation}
\begin{split}
	G^{(2)}_{2 \ TLS} (\vector{r}_{1},\vector{r}_{2})
	= 6 \ \mathcal{E}_{0}^{4} \left<\hat{n}\right>^{2}  \left(1 + \frac{1}{3} \cos(\delta_{1}-\delta_{2}) \right) \; .
\label{eq:G2TLS}
\end{split}
\end{equation}

In order to work with a probabilistic theory that remains valid (i.e., normalized) for arbitrary light sources, we
introduce in what follows two more detectors in the setup where the first pair of detectors is placed at arbitrary positions $\delta_{1}$ and $\delta_{2}$ and the second pair at positions $\pi_{1}\equiv \delta_{1}+\pi$ and $\pi_{2}\equiv \delta_{2}+\pi$.  With the four detectors at $\delta_{1}$, $\delta_{2}$, $\pi_{1}$ and $\pi_{2}$, it is possible to measure altogether six second order correlation functions.
 
In case of two SPE the six second order correlation functions $G^{(2)}_{2 \ SPE}$ read (cf. Eq. (\ref{eq:G2SPE}))
\begin{equation}
\begin{split}
\label{G2_2TLS}
	&G^{(2)}_{2 \ SPE}(\delta_{1},\delta_{2})=2 \ \mathcal{E}_{0}^{4}(1 + \mathcal{V}_{2 \ SPE} \cos(\delta_{1}-\delta_{2}))  \\
	&G^{(2)}_{2 \ SPE}(\pi_{1},\pi_{2})=2 \ \mathcal{E}_{0}^{4}(1 + \mathcal{V}_{2 \ SPE} \cos(\delta_{1}-\delta_{2}))  \\
	&G^{(2)}_{2 \ SPE}(\delta_{1},\pi_{2})=2 \ \mathcal{E}_{0}^{4}(1 - \mathcal{V}_{2 \ SPE} \cos(\delta_{1}-\delta_{2}))  \\
	&G^{(2)}_{2 \ SPE}(\pi_{1},\delta_{2})=2 \ \mathcal{E}_{0}^{4}(1 - \mathcal{V}_{2 \ SPE} \cos(\delta_{1}-\delta_{2}))  \\
	&G^{(2)}_{2 \ SPE}(\delta_{1},\pi_{1})=G^{(2)}_{2 \ SPE}(\delta_{2},\pi_{2})= 2 \ \mathcal{E}_{0}^{4}(1 - \mathcal{V}_{2 \ SPE}) \ ,
\end{split}
\end{equation}
where the visibility $\mathcal{V}_{2 \ SPE}$ has been added artificially to incorporate possible experimental imperfections \cite{Wiegner2010}. 
By summing over all six correlation functions we obtain the following normalization factor 
\begin{equation}
\label{norm_2TLS}
\mathcal{N}=  \sum_{i,j=1,2}G^{(2)}_{2 \ SPE}(\delta_{i},\pi_{j})+G^{(2)}_{2 \ SPE}(\delta_{1},\delta_{2})+ G^{(2)}_{2 \ SPE}(\pi_{1},\pi_{2})= 2 \, \mathcal{E}_{0}^{4} ( 6 - 2 \mathcal{V}_{2 \ SPE}  ) \ ,
\end{equation}
which can be used to define the second order correlation functions as probabilities. 
By use of Eqs.~(\ref{G2_2TLS}) and (\ref{norm_2TLS}) we thus obtain for the two-photon detection probabilities $P^{(2)}_{2 \ SPE}$ 
\begin{equation}
\begin{aligned}
\label{probab_SPE}
	&P^{(2)}_{2 \ SPE}(\delta_{1},\delta_{2})&=
	& \ \frac{1}{6 - 2 \mathcal{V}_{2 \ SPE}}(1+\mathcal{V}_{2 \ SPE}\cos(\delta_{1}-\delta_{2})) \\
	&P^{(2)}_{2 \ SPE}(\pi_{1},\pi_{2})&=
	& \ \frac{1}{6 - 2 \mathcal{V}_{2 \ SPE}}(1+\mathcal{V}_{2 \ SPE}\cos(\delta_{1}-\delta_{2}))  \\
	&P^{(2)}_{2 \ SPE}(\delta_{1},\pi_{2})&=
	& \ \frac{1}{6 - 2 \mathcal{V}_{2 \ SPE}}(1-\mathcal{V}_{2 \ SPE}\cos(\delta_{1}-\delta_{2}))  \\
	&P^{(2)}_{2 \ SPE}(\pi_{1},\delta_{2})&=
	& \ \frac{1}{6 - 2 \mathcal{V}_{2 \ SPE}}(1-\mathcal{V}_{2 \ SPE}\cos(\delta_{1}-\delta_{2}))  \\
	&P^{(2)}_{2 \ SPE}(\delta_{1},\pi_{1})&=& \ P^{(2)}_{2 \ SPE}(\delta_{2},\pi_{2})= \frac{1 - \mathcal{V}_{2 \ SPE}}{6 - 2 \mathcal{V}_{2 \ SPE}}  \ ,
\end{aligned}
\end{equation}
whereas the  probabilities for single photon detection events $P^{(1)}_{2 \ SPE}(\delta_{1})$ and $P^{(1)}_{2 \ SPE}(\delta_{2})$ are derived from the two-photon detection probabilities via
\begin{equation}
\begin{aligned}
\label{P(1)SPE}
	P^{(1)}_{2 \ SPE}(\delta_{1})=  & \sum_{\alpha = \pi_1, \pi_2, \delta_2} P^{(2)}_{2 \ SPE}(\delta_{1},\alpha) = \frac{3 - \mathcal{V}_{2 \ SPE}}{6 - 2 \mathcal{V}_{2 \ SPE}} = \ \frac{1}{2} \\
	P^{(1)}_{2 \ SPE}(\delta_{2})= & \sum_{\alpha = \pi_1, \pi_2, \delta_1} P^{(2)}_{2 \ SPE}(\alpha,\delta_{2}) = \frac{3 - \mathcal{V}_{2 \ SPE}}{6 - 2 \mathcal{V}_{2 \ SPE}} = \ \frac{1}{2} \; .
\end{aligned}
\end{equation}
Identifying in Eq.~(\ref{eq:ch74}) $x=P^{(1)}_{2 \ SPE}(\delta_{1})$, $x'=P^{(1)}_{2 \ SPE}(\delta_{1}')$, $y=P^{(1)}_{2 \ SPE}(\delta_{2})$ and $y'=P^{(1)}_{2 \ SPE}(\delta_{2}')$ as the probability to measure a single photon at position $\delta_{1}$, $\delta_{1}'$, $\delta_{2}$ and $\delta_{2}'$, respectively, and the products $xy, \ldots$ as the joint two-photon detection probabilities $P^{(2)}_{2 \ SPE}(\delta_{1},\delta_{2}), \ldots$, we arrive at the following set of Bell inequalities
\begin{equation}
\begin{split}
	& -1\leq  \ \frac{\mathcal{V}_{2 \ SPE}}{6 - 2 \mathcal{V}_{2 \ SPE}} \ \Big(\cos(\delta_{1}-\delta_{2})-\cos(\delta_{1}-\delta'_{2})\\
	&+\cos(\delta'_{1}-\delta_{2})+\cos(\delta'_{1}-\delta'_{2})\Big)+ \frac{2}{6 - 2 \mathcal{V}_{2 \ SPE}} - 1 \leq 0 \; .
\label{eq:ch74final}
\end{split}
\end{equation} 
Using for the arguments of the cosine-functions in Eq.~(\ref{eq:ch74final}) the upper bound Bell angles  $\pi/4$, $3\pi/4$, $\pi/4$, $\pi/4$ \cite{Bertlmann2006} for testing the upper bound, and the lower bound angles $3\pi/4$, $\pi/4$, $3\pi/4$, $3\pi/4$ for testing the lower bound, one obtains a violation of Eq.~(\ref{eq:ch74final}) for different visibilities $\mathcal{V}_{2\ SPE}$. Since the term in brackets obtains maximally (minimally) the value $2\sqrt{2}$ ($-2\sqrt{2}$) for the upper bound angles (lower bound angles) the upper bound can be violated for $\mathcal{V}_{2\ SPE} > 2 /(1+\sqrt{2}) \approx 0.83$ and the lower bound for $\mathcal{V}_{2\ SPE} > 1/\sqrt{2} \approx 0.71$. 
Note that for $\mathcal{V}_{2\ SPE} = 1$ we obtain from Eqs.~(\ref{probab_SPE}) and (\ref{P(1)SPE}) for the two-photon detection probabilities $P^{(2)}_{2 \ SPE}$ 
\begin{equation}
\begin{aligned}
	&P^{(2)}_{2 \ SPE}(\delta_{1},\delta_{2})&=
	& \ \frac{1}{4}(1+\cos(\delta_{1}-\delta_{2})) \\
	&P^{(2)}_{2 \ SPE}(\pi_{1},\pi_{2})&=
	& \ \frac{1}{4}(1+\cos(\delta_{1}-\delta_{2}))  \\
	&P^{(2)}_{2 \ SPE}(\delta_{1},\pi_{2})&=
	& \ \frac{1}{4}(1-\cos(\delta_{1}-\delta_{2}))  \\
	&P^{(2)}_{2 \ SPE}(\pi_{1},\delta_{2})&=
	& \ \frac{1}{4}(1-\cos(\delta_{1}-\delta_{2}))  \\
	&P^{(2)}_{2 \ SPE}(\delta_{1},\pi_{1})&=& \ P^{(2)}_{2 \ SPE}(\delta_{2},\pi_{2})= 0  \ ,
\end{aligned}
\end{equation}
and for the single photon detection probabilities 
\begin{equation}
\begin{aligned}
	P^{(1)}_{2 \ SPE}(\delta_{1})=  & \sum_{\alpha = \pi_1, \pi_2, \delta_2} P^{(2)}_{2 \ SPE}(\delta_{1},\alpha) = \ \frac{1}{2} \\
	P^{(1)}_{2 \ SPE}(\delta_{2})= & \sum_{\alpha = \pi_1, \pi_2, \delta_1} P^{(2)}_{2 \ SPE}(\alpha,\delta_{2}) = \ \frac{1}{2} \; ,
\end{aligned}
\end{equation}
so that we are left with the more familiar Bell inequality \cite{Clauser74}
\begin{equation}
\label{Bellfamiliar}
	-1\leq \ \frac{1}{4} \  \Big(\cos(\delta_{1}-\delta_{2})-\cos(\delta_{1}-\delta'_{2})
	+\cos(\delta'_{1}-\delta_{2})+\cos(\delta'_{1}-\delta'_{2})\Big)-\frac{1}{2} \leq 0 \; .
\end{equation}

In what follows we employ the same approach as above to study whether a violation of the Bell inequalities Eq.~(\ref{eq:ch74}) can be obtained also for photons emitted by two statistically independent TLS.
To this goal we derive by use of Eq. (\ref{eq:G2TLS}) the following set of second order correlation functions
\begin{equation}
\begin{aligned}
\label{g2TLS}
	&G_{2 \ TLS}^{(2)}(\delta_{1},\delta_{2})&=& \ 6 \ \mathcal{E}_{0}^{4} \left<\hat{n}\right>^{2}  \left( 1+\mathcal{V}_{2 \ TLS}\cos(\delta_{1}-\delta_{2}) \right)\\
	&G_{2 \ TLS}^{(2)}(\pi_{1},\pi_{2})&=& \ 6 \ \mathcal{E}_{0}^{4} \left<\hat{n}\right>^{2}  \left( 1+\mathcal{V}_{2 \ TLS}\cos(\delta_{1}-\delta_{2}) \right) \\
	&G_{2 \ TLS}^{(2)}(\delta_{1},\pi_{2})&=& \ 6 \ \mathcal{E}_{0}^{4} \left<\hat{n}\right>^{2}  \left( 1-\mathcal{V}_{2 \ TLS}\cos(\delta_{1}-\delta_{2}) \right) \\
	&G_{2 \ TLS}^{(2)}(\pi_{1},\delta_{2})&=& \ 6 \ \mathcal{E}_{0}^{4} \left<\hat{n}\right>^{2}  \left( 1-\mathcal{V}_{2 \ TLS}\cos(\delta_{1}-\delta_{2}) \right) \\
	&G_{2 \ TLS}^{(2)}(\delta_{1},\pi_{1})&=& \ G_{2 \ TLS}^{(2)}(\delta_{2},\pi_{2}) =   \ 6 \ \mathcal{E}_{0}^{4} \left<\hat{n}\right>^{2} \left( 1-\mathcal{V}_{2 \ TLS} \right) \, ,
\end{aligned}
\end{equation} 
where, in analogy with  Eq.~(\ref{G2_2TLS}), we introduced again artificially the visibility $\mathcal{V}_{2 \ TLS}$ to incorporate experimental imperfections and also to infer the least visibility required to violate the Bell inequalities.
Again, the sum over all second order correlation functions yields the normalization factor transforming the correlation functions into probabilities, which this time calculates to 
\begin{equation}
\label{norm_2TLS_1}
\mathcal{N}=  \sum_{i,j=1,2}G^{(2)}_{2 \ TLS}(\delta_{i},\pi_{j})+G^{(2)}_{2 \ TLS}(\delta_{1},\delta_{2})+ G^{(2)}_{2 \ TLS}(\pi_{1},\pi_{2})=  6 \ \mathcal{E}_{0}^{4} \left<\hat{n}\right>^{2} \left( 6-2\mathcal{V}_{2 \ TLS} \right) \ .
\end{equation}
In analogy to Eqs.~(\ref{probab_SPE}) and~(\ref{P(1)SPE}) this allows to derive the following  joint two-photon and single photon detection probabilities
\begin{equation}
\begin{split}
\label{probabTLS_1}
	P^{(2)}_{2 \ TLS}(\delta_{1},\delta_{2})= & \frac{1}{6-2\mathcal{V}_{2 \ TLS}}(1+\mathcal{V}_{2 \ TLS}\cos(\delta_{1}-\delta_{2})) \\
	P^{(2)}_{2 \ TLS}(\pi_{1},\pi_{2})= & \frac{1}{6-2\mathcal{V}_{2 \ TLS}}(1+\mathcal{V}_{2 \ TLS}\cos(\delta_{1}-\delta_{2})) \\
	P^{(2)}_{2 \ TLS}(\delta_{1},\pi_{2})= & \frac{1}{6-2\mathcal{V}_{2 \ TLS}}(1-\mathcal{V}_{2 \ TLS}\cos(\delta_{1}-\delta_{2})) \\
	P^{(2)}_{2 \ TLS}(\pi_{1},\delta_{2})= & \frac{1}{6-2\mathcal{V}_{2 \ TLS}}(1-\mathcal{V}_{2 \ TLS}\cos(\delta_{1}-\delta_{2})) \\ 
	P^{(2)}_{2 \ TLS}(\delta_{1}, \pi_{1})= & P^{(2)}_{2 \ TLS}(\delta_{2}, \pi_{2})=\frac{1-\mathcal{V}_{2 \ TLS}}{6-2 \mathcal{V}_2 \ TLS} \; ,
\end{split}
\end{equation}
and
\begin{equation}
\label{probabTLS_2}
	P^{(1)}_{2 \ TLS}(\delta_{1})= P^{(1)}_{2 \ TLS}(\delta_{2})=\frac{3-\mathcal{V}_{2 \ TLS}}{6-2\mathcal{V}_{2 \ TLS}}=\frac{1}{2} \; .
\end{equation}
Plugging $P^{(2)}_{2 \ TLS}(\delta_1, \delta_2)$ from Eq.~(\ref{probabTLS_1}) and $P^{(1)}_{2 \ TLS}(\delta_{1})$, $P^{(1)}_{2 \ TLS}(\delta_{2})$ from Eq.~(\ref{probabTLS_2}) into Eq.~(\ref{eq:ch74}) we obtain 
\begin{equation}
\begin{split}
&	-1\leq  \ \frac{\mathcal{V}_{2 \ TLS}}{6-2\mathcal{V}_{2 \ TLS}} \ \Big(\cos(\delta_{1}-\delta_{2})-\cos(\delta_{1}-\delta'_{2})\\
&	+\cos(\delta'_{1}-\delta_{2})+\cos(\delta'_{1}-\delta'_{2})\Big)+\frac{2}{6-2\mathcal{V}_{2 \ TLS}}-1 \leq 0 \, .
\label{eq:ch74TLSVis}
\end{split}
\end{equation}
Since Eq.~(\ref{eq:ch74TLSVis}) is structurally identical to Eq.~(\ref{eq:ch74final}), the upper and lower bound can be be violated for identical visibilities as in Eq.~(\ref{eq:ch74final}), i.e., for $\mathcal{V}_{2\ TLS} > 2 /(1+\sqrt{2}) \approx 0.83$ for the upper bound and $\mathcal{V}_{2\ TLS} > 1/\sqrt{2} \approx 0.71$ for the lower bound.
However, the visibility of the second order correlation function for two photons emitted from two statistically independent classical sources acquires at most a value $\mathcal{V}=1/2$, which is the case for two statistically independent coherent sources ($\mathcal{V}=1/3$ for two statistically independent TLS), even in the ideal case of negligible experimental insufficiencies \cite{Mandel83}. It is obvious that with this value for the visibilities the Bell inequalities of Eq.~(\ref{eq:ch74TLSVis}) can not be violated. 

\begin{figure}[h]
	\centering
\includegraphics[width=0.7\textwidth]{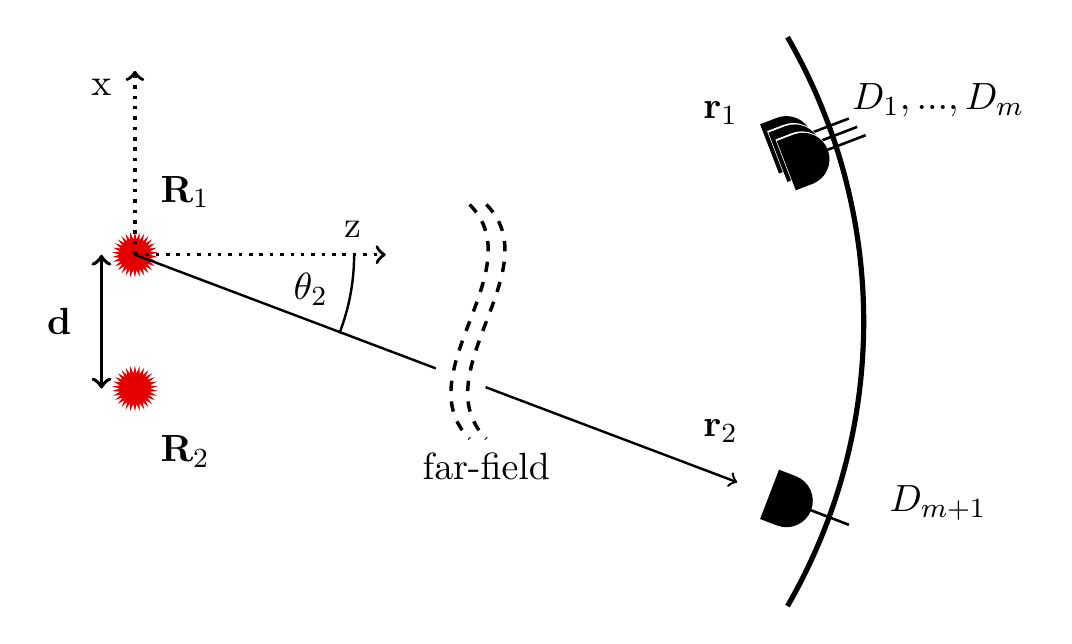}
\caption{Scheme of the modified setup. In contrast to Fig.~1, $m$ detectors $D_i$, $i = 1, \ldots, m$, are located in the far field at $\vector{r}_{1}$ and one detector $D_{m+1}$ at $\vector{r}_{2}$. Assuming that each detector registers coincidentally exactly one photon the $(m+1)$-th order correlation function $G^{(m+1)}_{2 \ TLS}(\vector{r}_{1}, \vector{r}_{2})$ is measured.}
\label{fig:detectionschemePlus1}
\end{figure}

\section{Violation of Bell's inequalities using higher order intensity correlations}
In this section we investigate whether by recording $m+1$ photons ($m \geq 2$) emitted from two statistically independent TLS 
we can produce a visibility $\mathcal{V}^{(m+1)}_{2 \ TLS}$ high enough to violate Eq.~(\ref{eq:ch74}).  Hereby, we assign via post-selection valid detection events to a simultaneous detection of $m$ photons at $\delta_{1}$ (or $\pi_{1}$) and one photon at $\delta_{2}$ (or $\pi_{2}$), whereas simultaneous detection events of $m$ photons at $\delta_{1}$ \textit{and} $m$ photons at $\pi_{1}$ or $1$ photon at $\delta_{2}$ \textit{and} $1$ photon at $\pi_{2}$ are not taken into account. The corresponding setup is displayed in Fig.~2. For this configuration the  $(m+1)$-th order correlation functions 
reads \cite{Shih09}
\begin{equation}
\begin{split}
\label{g(m+1)2TLS}
	G^{(m+1)}_{2 \ TLS}(\delta_{1},\delta_{2})= & \mathcal{A} \ (1+\mathcal{V}^{(m+1)}_{2 \ TLS}\cos(\delta_{1}-\delta_{2}))\\
	G^{(m+1)}_{2 \ TLS}(\pi_{1},\pi_{2})= & \mathcal{A} \ (1+\mathcal{V}^{(m+1)}_{2 \ TLS}\cos(\delta_{1}-\delta_{2})) \\
	G^{(m+1)}_{2 \ TLS}(\delta_{1},\pi_{2})= & \mathcal{A} \ (1-\mathcal{V}^{(m+1)}_{2 \ TLS}\cos(\delta_{1}-\delta_{2})) \\
	G^{(m+1)}_{2 \ TLS}(\pi_{1},\delta_{2})= & \mathcal{A} \ (1-\mathcal{V}^{(m+1)}_{2 \ TLS}\cos(\delta_{1}-\delta_{2})) \; ,
\end{split}
\end{equation}
where the visibility $\mathcal{V}^{(m+1)}_{2 \ TLS}$ is given by $\mathcal{V}^{(m+1)}_{2 \ TLS}=\frac{m}{m+2}$ and $\mathcal{A}$ is a constant given by $\mathcal{A} = \frac{(m+2)!}{(m+1)} 2^{m} \mathcal{E}_{0}^{2(m+1)} \left<\hat{n}\right>^{m+1}$. 
Again, summing over all second order correlation functions yields the normalization factor 
\begin{equation}
\label{Norm:m+1TLS}
\mathcal{N}=\sum_{\substack{i=\delta_{1},\pi_{1}\\ j=\delta_{2},\pi_{2}}}G^{(m+1)}_{2 \ TLS}(i,j)= 4 \ \mathcal{A} \ ,
\end{equation}
so that the corresponding joint two-photon and single photon detection probabilities, derived from Eqs.~(\ref{g(m+1)2TLS}) and (\ref{Norm:m+1TLS}), read
\begin{equation}
\begin{split}
	&P^{(2)}_{2 \ TLS}(\delta_{1},\delta_{2})= \ \frac{1}{4}(1+\mathcal{V}^{(m+1)}_{2 \ TLS}\cos(\delta_{1}-\delta_{2})) \\
	&P^{(2)}_{2 \ TLS}(\pi_{1},\pi_{2})=\frac{1}{4}(1+\mathcal{V}^{(m+1)}_{2 \ TLS}\cos(\delta_{1}-\delta_{2})) \\
	&P^{(2)}_{2 \ TLS}(\delta_{1},\pi_{2})=\frac{1}{4}(1-\mathcal{V}^{(m+1)}_{2 \ TLS}\cos(\delta_{1}-\delta_{2})) \\
	&P^{(2)}_{2 \ TLS}(\pi_{1},\delta_{2})=\frac{1}{4}(1-\mathcal{V}^{(m+1)}_{2 \ TLS}\cos(\delta_{1}-\delta_{2})) \; ,
\end{split}
\end{equation}
and 
	\begin{equation}
	P^{(1)}_{2 \ TLS}(\delta_{1})=
	P^{(1)}_{2 \ TLS}(\delta_{2})=
	\ \frac{1}{2} \ .
\end{equation}
In this case the Bell inequalities Eq.~(\ref{eq:ch74}) take the following form
\begin{equation}
	-1\leq \ \frac{1}{4} \ \mathcal{V}^{(m+1)}_{2 \ TLS}\ \Big(\cos(\delta_{1}-\delta_{2})-\cos(\delta_{1}-\delta'_{2})
	 +\cos(\delta'_{1}-\delta_{2})+\cos(\delta'_{1}-\delta'_{2})\Big)-\frac{1}{2} \leq 0 \; .
\label{eq:ch74TLSNplus1}
\end{equation}

It can easily be seen that, after replacing $\mathcal{V}^{(m+1)}_{2 \ TLS}$ by $\mathcal{V}_{2 \ SPE}$ and assuming $\mathcal{V}_{2 \ SPE} = 1$, this equation matches with Eq.~(\ref{Bellfamiliar}) for two SPE. Moreover, a violation of Eq.~(\ref{eq:ch74TLSNplus1}) can be achieved if $\mathcal{V}^{(m+1)}_{2 \ TLS}=\frac{m}{m+2} > 1/\sqrt{2}$. This is realized for $m \geq 5$ in which case we have $\mathcal{V}^{(5+1)}_{2 \ TLS} \approx 0.714>1/\sqrt{2}$. 
This violation of Bell's inequalities demonstrates that measuring higher order correlation functions can indeed create and herald non-local multiparticle entanglement among photons. The interpretation for the considered setup is that a bunch of $m \geq 5$ photons at position $\delta_{1}$ is entangled with one photon at $\delta_{2}$, all $m+1$ photons being emitted by classical light sources, in this case by two statistically independent TLS.  

\begin{figure}[h!]
	\centering
		\includegraphics[width=0.8\textwidth]{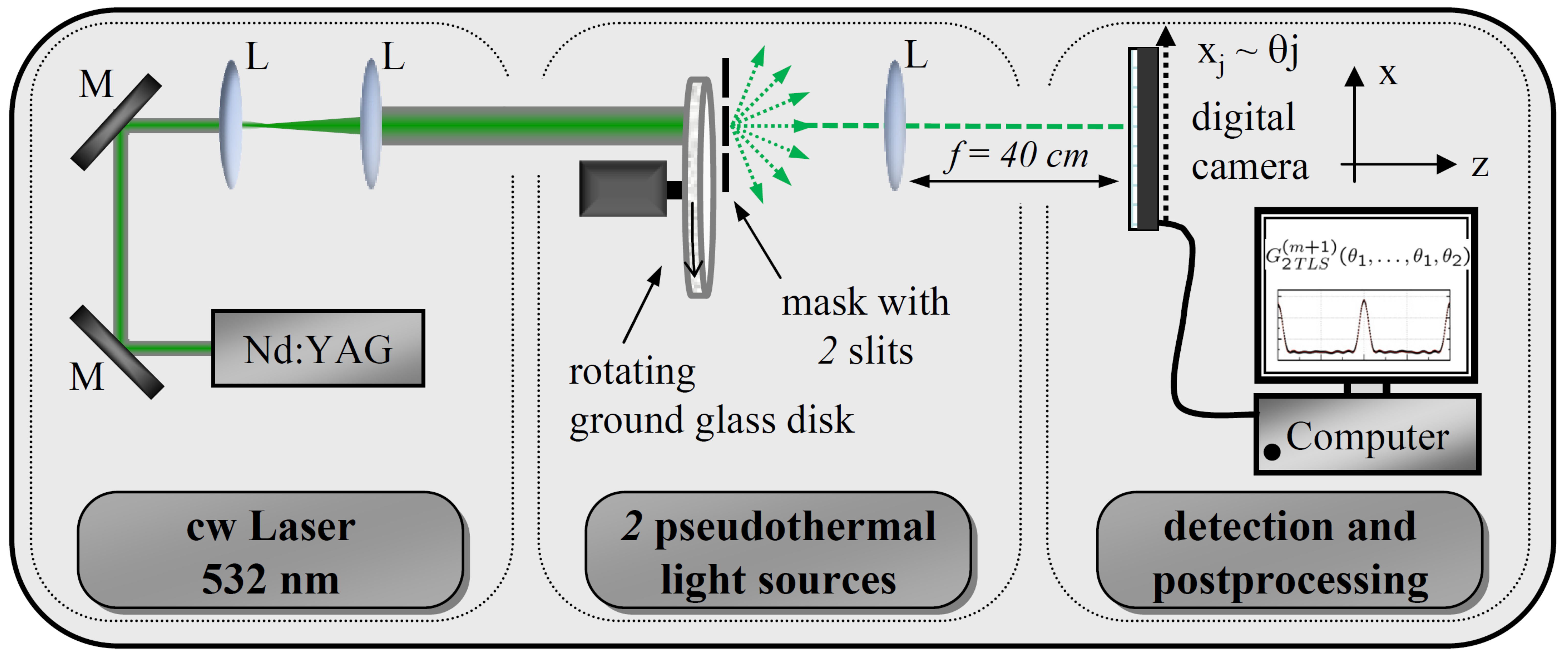}
	\caption{Experimental setup to measure the $(m+1)$-th order correlation function $G^{(m+1)}_{2 \ TLS}(\delta_{1},\delta_{2})$ with $2$ Thermal Light Sources (TLS). For details see text. M: mirror; L: lens.}
	\label{fig:Fig3_Zanthier}
\end{figure}

The entanglement of the photons can be understood also from a different perspective. The detection of the first $m$ photons can be interpreted as a projection of the two TLS sources onto a state which emits the subsequent photon not uniformly but in a strongly correlated manner, such that the Bell inequalities are violated. 
This projection of the sources can be seen as similar to Einstein Podolsky Rosen steering by photon subtraction as recently discussed in \cite{Agarwal2014}.
From this point of view the entanglement and violation of Bell's inequalities is a consequence of the nonvanishing cross correlations between noncommuting quadrature phase components of the two spatially separated fields of the type  $\langle \hat{a}^{\dagger}_{0} \, \hat{a}_{1} \rangle_{\rho^{(m)}}$, with $\langle \hat{a}^{\dagger}_{0} \rangle_{\rho^{(m)}} = \langle \hat{a}_{1} \rangle_{\rho^{(m)}} =0$, where $\rho^{(m)}$ is the density matrix of the field after $m$ photons have been recorded. 
Indeed, for the cross correlation coefficient $C^{(m)} = \langle \hat{a}^{\dagger}_{0} \hat{a}_{1} \rangle_{\rho^{(m)}}/\sqrt{\langle \hat{a}^{\dagger}_{0} \hat{a}_{0} \rangle_{\rho^{(m)}} \langle \hat{a}^{\dagger}_{1} \hat{a}_{1} \rangle_{\rho^{(m)}}}$ \cite{Agarwal2014} we obtain $C^{(m)} =\frac{m}{m+2} = \mathcal{V}^{(m+1)}_{2 TLS}$, which approaches unity in the limit $m \rightarrow \infty$.

\section{Experimental setup and results}
To measure $G^{(m+1)}_{2 \ TLS}(\delta_{1},\delta_{2})$ with two statistically independent TLS we used a mask with two identical slits of width $a = 25 \, \mu$m and separation $d = 200 \, \mu$m. The mask is located a few centimeters behind a rotating ground glass disk which is illuminated by a linearly polarized frequency-doubled Nd:YAG laser at $\lambda = 532$ nm (see Fig.~3).  
Within each slit a large number of time-dependent speckles is produced, generated by the stochastically interfering waves scattered from the granular surface of the ground glass disk, representing many independent point-like sub-sources equivalent to an ordinary spatially incoherent thermal source. The coherence time of the pseudothermal sources depends on the rotational speed of the disk \cite{Tuft71} and was chosen in the experiment to $\tau_c \approx 50$ ms. To ensure that the two TLS radiate with equal intensity we enlarged the incident laser beam by a microscope objective to 1~cm providing a homogeneous illumination of the mask. 
To determine $G^{(m+1)}_{2 \ TLS}(\delta_{1},\delta_{2})$ we measured the intensity of the light scattered by the two TLS at different positions using a conventional digital camera placed in the focal point (Fourier plane) of a lens behind the mask ($z \approx f$) thus fulfilling the far field condition (see Fig.~3). 
Hereby, each pixel of the camera serves as a detector to register the intensity at position $x_j/z \sim \theta_j$, and we correlated $m-1$ pixels at $x_1 \sim \theta_1$, each separated by one pixel along the $y$-direction to use $m-1$ different pixels with identical $x_1$-values, with another pixel at $x_2 \sim \theta_2$. 
With more than a million of pixels the digital camera has the advantage that the amount of data accumulated in one frame to correlate the intensities at $m$ different pixels is exceedingly larger than using $m$ single photon detectors \cite{Oppel2012}. In order to obtain interference signals of high visibility, the integration time of the camera $\tau_{i}$ was chosen much shorter than the coherence time of the TLS, in our case $\tau_{i} \approx 3$~ms $<< \tau_{c}$. 

Fig.~4 displays the experimentally observed visibilities of the normalized $(m+1)$-th order correlation functions $\mathrm{g}^{(m+1)}_{2 \ TLS}(\vector{r}_{1}, \vector{r}_{2}) = \mathrm{G}^{(m+1)}_{2 \ TLS}(\vector{r}_{1}, \vector{r}_{2})/((\mathrm{G}^{(1)}_{2 \ TLS} (\vector{r}_{1}))^{m} \mathrm{G}^{(1)}_{2 \ TLS} (\vector{r}_{2}))$ for $m + 1 = 2, \ldots, 9$. For illustration, the measured normalized correlation function $g^{(6+1)}_{2 \ TLS}(\vector{r}_{1}, \vector{r}_{2})$  as a function of the position of detector $D_{m+1}$ at $\vector{r}_{2}$ (for $\vector{r}_{1}=0$) is shown in the inset of Fig.~4 and compared to the theoretical prediction of Eq.~(\ref{g(m+1)2TLS}).
It can be seen that the experimental results are in excellent agreement with the theory, in particular that the experimentally obtained visibilities follow the expression $\mathcal{V}^{(m+1)}_{2 \ TLS}=\frac{m}{m+2}$.
However, one can also see that experimentally for $m=5$, due to remaining experimental uncertainties, a violation of Bell's inequalities is not yet obtained, in contrast to the theoretical prediction $\mathcal{V}^{(5+1)}_{2 \ TLS}=\frac{5}{5+2} = 0.714 > 1/\sqrt{2}$. Nonetheless, a visibility of $\mathcal{V}^{(6+1)}_{2 \ TLS} = 0.743 \pm 0.027$ is achieved experimentally for $m=6$. According to Eq.~(\ref{eq:ch74TLSNplus1}) this visibility demonstrates a violation of the Bell inequalities.

\begin{figure}
	\centering
		\includegraphics[width=0.7\textwidth]{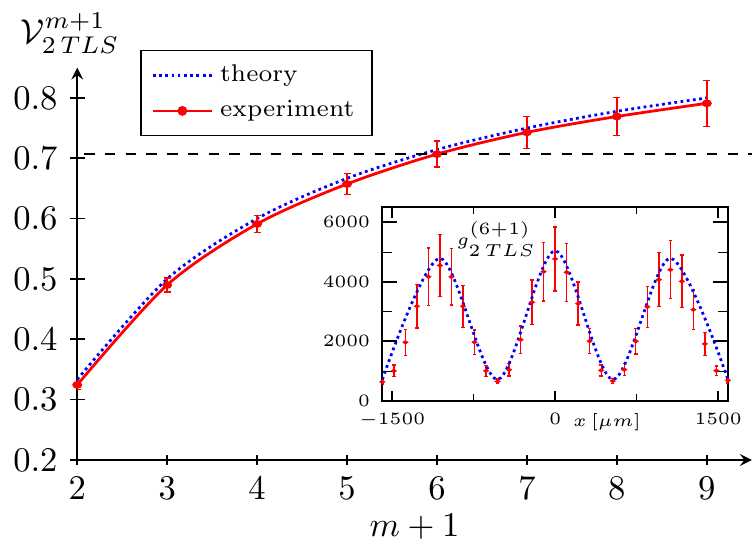}		
	\caption{Visibilities $\mathcal{V}^{(m+1)}_{2 \ TLS}$ of the normalized $(m+1)$-th order correlation functions $\mathrm{g}^{(m+1)}_{2 \ TLS}(\vector{r}_{1}, \vector{r}_{2}) = \mathrm{G}^{(m+1)}_{2 \ TLS}(\vector{r}_{1}, \vector{r}_{2})/((\mathrm{G}^{(1)}_{2 \ TLS} (\vector{r}_{1}))^{m} \mathrm{G}^{(1)}_{2 \ TLS} (\vector{r}_{2}))$ for two Thermal Light Sources (TLS) for $m + 1 = 2, \ldots, 9$. The blue (dotted) curve shows the theoretical prediction $\mathcal{V}^{(m+1)}_{2 \ TLS} = \frac{m}{m+2}$, the red (solid) curve the experimental results. In the inset is shown the normalized correlation function $g^{(6+1)}_{2 \ TLS}(\vector{r}_{1}, \vector{r}_{2})$ for $m + 1 = 7$ as a function of the position of the detector at $\vector{r}_{2}$ for $\vector{r}_{1}=0$ (blue (dotted) curve: theory; red points: experimental results). The error bars reflect the statistical uncertainties.}
	\label{fig:Fig4_Zanthier}
\end{figure}

\section{Conclusion}
In conclusion we showed that spatially separated photons emitted by statistically independent classical sources may display spatial correlations strong enough to violate Bell's inequalities. In contrast to SPE \cite{Wiegner2010}, the violation does not occur for two recorded photons but for $m \geq 5$ photons registered in one mode and one photon in another mode. Due to experimental uncertainties the violation was demonstrated experimentally for $m \geq 6$. The spatial correlations among the photons can be understood from the perspective of state projection where the detection of the first $m$ photons projects the sources onto a state which emits the subsequent photon in a strongly correlated manner such that Bell's inequalities are violated.  From this perspective the observed entanglement and violation of Bell's inequalities is due to cross correlations among noncommuting quadrature phase components of the two spatially separated fields only appearing after $m$ photons have been recorded. 

\section{Acknowledgment}
The authors thank G. S. Agarwal for fruitful discussions. R. Schneider, T. Mehringer, S. Oppel and J. von Zanthier thank the Erlangen Graduate School in Advanced Optical Technologies (SAOT) by the German Research Foundation (DFG) in the framework of the German excellence initiative for funding. This work was supported by the DFG under grant ZA 293/4-1.

\end{document}